\documentclass[a4paper,12pt,english]{article}

\usepackage{amsfonts,bm,amssymb,euscript,array,babel,cite}
\usepackage[all,color]{xy}
\usepackage{relsize}
\usepackage{tikz}
\usepackage{url}
\usepackage{amsmath,amsthm,accents}
\usepackage[]{graphicx}
\usepackage[T1]{fontenc}
\usepackage{framed}

\usepackage{empheq}

\makeatletter

\newcommand{\co}{{\cal O}}

\newcommand{\dd}{\mathrm{d}}
\newcommand{\dD}{\mathrm{D}}

\newcommand{\w}{\wedge}
\newcommand{\bbm}{\left(\begin{matrix}}
\newcommand{\ebm}{\end{matrix}\right)}
\newcommand{\beq}{\begin{eqnarray}}
\newcommand{\eeq}{\end{eqnarray}}

\newcommand{\eqn}[1]{\begin{equation}
                        \begin{split} #1 \end{split}
                    \end{equation}}
\newcommand{\half}{\frac{1}{2}}

\makeatother

\newtheorem{prop}{Proposition}[section]

\newtheorem{obser}{Observation}[section]

\newcommand{\sfrac}[2]{{\textstyle\frac{#1}{#2}}}

\newcommand{\be}{\begin{equation}}
\newcommand{\ee}{\end{equation}}

\newcommand{\beqa}{\begin{eqnarray}}
\newcommand{\eeqa}{\end{eqnarray}} 
\def\nn{\nonumber} \def \bea{\begin{eqnarray}} \def\eea{\end{eqnarray}}

\newcommand{\barr}{\begin{array}}
\newcommand{\earr}{\end{array}}
\numberwithin{equation}{section}
\newcommand{\mf}{\mathfrak}

\def\a{\alpha}  
  \def\G{\Gamma}
 \def\d{\delta} \def\D{\Delta}

   \def\m{\mu}
 \def\o{\omega} \def\p{\pi}  
\def\s{\sigma} \def\S{\Sigma}

 \def\one{\mbox{1 \kern-.59em {\rm l}}}

\def\bit{\begin{itemize}} \def\eit{\end{itemize}}

\def\({\left(} \def\){\right)}

 \allowdisplaybreaks[3]

\begin{document}

\makeatother

%%%%%%%%%%%%%%%%%%%%%%%%%%%%%%%%%%%%%%%%%%%%%%%%%%%%%%%%%%%%%%%%%%%%%%%%%%%%%%%%
%%%%%%%%%

\parindent=0cm

\renewcommand{\title}[1]{\vspace{10mm}\noindent{\Large{\bf

#1}}\vspace{8mm}} \newcommand{\authors}[1]{\noindent{\large

#1}\vspace{5mm}} \newcommand{\address}[1]{{\itshape #1\vspace{2mm}}}

%\thispagestyle{hepth}

%%%% --- TITLE PAGE --- %%%%
\begin{titlepage}

% \begin{flushright}
% \today \\
% \end{flushright}

\begin{center}

%\vskip 3mm

\title{ {\Large 
Beyond the standard gauging: \\ \vspace{3pt} gauge symmetries of Dirac Sigma Models
} }

 \authors{ Athanasios {Chatzistavrakidis}$\footnote{a.chatzistavrakidis@rug.nl}$, Andreas Deser$\footnote{andreas.deser@itp.uni-hannover.de}$,\\
 %$~^{\sharp,}$
 Larisa Jonke$\footnote{larisa@irb.hr}$
 %$~^{\dagger,}$ 
 and Thomas Strobl$\footnote{{stroblATmath.univ-lyon1.fr}}$

 }

\vskip 1mm

\address{
$^{1}$Van Swinderen Institute for Particle Physics and 
Gravity, University of Groningen, \\ 
Nijenborgh 4, 9747 AG Groningen, The Netherlands

$^{2}$
 Institut f{\"u}r Theoretische Physik,
 Leibniz Universit{\"a}t Hannover, \\
 Appelstr.2, 30167 Hannover, Germany

$^{3}$ 
 Division of Theoretical Physics, 
 Rudjer Bo$\check s$kovi\'c Institute, \\
 Bijeni$\check c$ka 54, 10000  Zagreb, Croatia

$^{4}$
Institut Camille Jordan, Universit\'e Claude Bernard Lyon 1, \\
43 boulevard du 11 novembre 1918, 69622 Villeurbanne cedex, France

}

\smallskip

\end{center}

\vskip 1mm

 \begin{center}
%%%% --- ABSTRACT --- %%%%
\textbf{Abstract}
\vskip 3mm
\begin{minipage}{14cm}%
In this paper we study the general conditions that have to be met for a gauged extension of a two-dimensional bosonic $\sigma$-model to exist.
 In an inversion of the usual approach of identifying a global symmetry and then promoting it to a 
local one, we focus directly on the gauge symmetries of the theory. This allows for action functionals 
which are gauge invariant for rather general background fields in the sense that their invariance conditions 
 are milder than the usual case. In particular, the vector fields that control 
the gauging need not be Killing. The relaxation of isometry for the background fields is controlled by 
two connections on a Lie algebroid $L$ in which the gauge fields take values, in a generalization of the common 
Lie-algebraic picture. Here we show that these connections can always be determined 
when $L$ is a Dirac structure in the $H$-twisted Courant algebroid. 
This also leads us to a derivation of the general form for the gauge symmetries of a wide class of two-dimensional topological field theories called Dirac $\sigma$-models, which interpolate 
between the G/G Wess-Zumino-Witten model and the (Wess-Zumino-term twisted) Poisson sigma model.
\end{minipage}

 \end{center}

\end{titlepage}

\tableofcontents

\section{Introduction}

%---Gauge symmetries

When does a problem, for example a variational one with some associated action functional, have a local symmetry? 
This is an interesting question due to the importance of local symmetries in nearly every corner of
theoretical physics.
Gauge theories exhibiting some local symmetry are usually related to 
the introduction of spacetime 1-forms (gauge fields) taking values in some Lie algebra $\mathfrak{g}$. 
This procedure has as starting point the identification of some global symmetry of the theory which is 
then promoted to a local one with the aid of these gauge fields. However, from a mathematical perspective, 
 symmetries are associated to more general structures than groups and algebras. These structures are 
 groupoids and algebroids (see the inspiring Ref. \cite{weinsteinams}).
The purpose of the present paper is to study some aspects of the role of Lie algebroid 
 structures in the simple physical setting of two-dimensional bosonic $\sigma$-models.
 To this end we first present some material from our upcoming, more mathematically oriented work \cite{cdjs}, 
 as a basis and motivation for the new results presented in this paper.

%---2D sigma models

Two-dimensional bosonic $\sigma$-models are based on maps $X=(X^i):\Sigma\to M, i=1,\dots,\text{dim}M$ from 
a two-dimensional source space $\Sigma$ of Lorentzian or Euclidean signature to a target space 
$M$. These theories are interesting because they describe strings propagating in target spacetimes $M$ through the scalar fields $X^i$.{\footnote{ For the purposes of the present paper we are not going to consider fermions and supersymmetry but rather focus on the classical bosonic theory.}} 
The source space is equipped with a metric $h=(h_{\a\beta}), \a,\beta=0,1$, which is a dynamical field in the theory. 
Furthermore, since the massless NSNS spectrum of string theory contains a Riemannian metric $g=(g_{ij})$ on the target space $M$, an antisymmetric Kalb-Ramond 2-form $B=(B_{ij})$ and a scalar dilaton $\Phi$, the relevant bosonic $\sigma$-model is 
\be \label{string}
S[X,h]=\frac 1{4\pi\a'}\int_{\S}\sfrac 12 g_{ij}(X)\dd X^i\w\ast\dd X^j+
\frac 1{4\pi\a'}\int_{\S}\sfrac 12 B_{ij}(X)\dd X^i\w \dd X^j + \frac 1{8\pi}\int_{\S}R\Phi\ast 1~,
\ee   
where $R$ is the world sheet curvature scalar.{\footnote{Since the dilaton coupling contributes at one-loop 
rather than at leading order it will be ignored in most parts of the analysis, which will 
remain classical in this paper.}}
This action depends on the world sheet metric $h$ through the Hodge star operator $\ast$, with 
$\ast^2=\mp 1$ for Euclidean and Lorentzian signature, respectively. Our first purpose then 
is to find a gauged extension of this action functional for as general background fields $g(X)$ and 
$B(X)$ as possible. 

%---Gauging 2D sigma models

The motivation to consider gauged versions of the $\sigma$-model is twofold. The first 
motivation is the intention to derive new theories on 
quotient spaces \cite{cdjs}, similarly to the case of gauging the Wess-Zumino-Witten (WZW) model whose target space is a 
group $G$ \cite{wzw}. When the gauging is along (the adjoint action of) a subgroup $H$ of $G$ one obtains the $G/H$ WZW models, $G/G$ being an 
extremal case \cite{gk1,gk2}. Another motivation to consider gauged $\sigma$-models is the study of target space duality in string theory. Recall that the celebrated Buscher rules relating two dual string backgrounds with Abelian isometries 
are derived by 
considering an intermediate gauge theory which reduces to the two dual actions upon integrating out 
different sets of fields, be they additional scalar fields (Lagrange multipliers) or the (non-dynamical) gauge 
fields that are coupled to the theory \cite{Buscher,Rocek}. A similar procedure may be followed in the case of non-Abelian 
isometries \cite{hs1,hs2,Alvarez1,Alvarez2}. 

An important difference of our procedure with respect to the traditional 
gauging is that we follow an inverted logic. Instead of identifying global symmetries and 
then promoting them to local ones, we are looking for more general gauge invariant extensions of 
the action functional \eqref{string} for arbitrary background fields.  The crucial difference is then that 
the invariance conditions for the background fields imposed in the case of a global symmetry are 
relaxed and replaced by much milder ones. This point of view was originally introduced in \cite{ks1} 
and studied further in \cite{kss,ks2,ks3,cdj}.

%---Our threefold generalization

The above procedure amounts to a threefold generalization of the standard picture of gauged $\sigma$-models 
with $\mf{g}$-valued gauge fields, minimally-coupled to the scalar degrees of freedom $X^i$ and 
with strong invariance conditions on the background fields. The first generalization is that we allow the gauge fields to take values in some Lie algebroid $L$ instead of a Lie algebra.\footnote{This idea goes back to \cite{Bojowald-Kotov-Strobl,StroblPRL} and, as seen only recently \cite{cdjs}, can be generalized to just anchored bundles covering a foliation.} 

%---1st fold
A clear and physically motivated way to introduce this structure goes through the well-known Cartan problem. Following the lecture notes \cite{crainic}, in this problem 
we consider two functions $C^a_{bc}$ and $e^i_a$ defined on an open set $U\subset \mathbb{R}^n$, with $a=1,\dots, r$ and $i=1,\dots n$. Then the aim of the problem is to 
determine (i) a manifold $N$, (ii) a set of 1-forms $\{e^a\}$ being a coframe on $N$, 
and (iii) a map $\varphi: N\to U\subset \mathbb{R}^n$, such that the following two equations hold:
\bea 
\dd e^a&=&-\sfrac 12 C^a_{bc}(\varphi)e^b\w e^c~,\label{mc}
\\
\dd \varphi^i&=&e^i_{a}(\varphi)e^a~.
\eea
Taking the exterior derivative on each side of each equation we directly obtain two necessary conditions, 
\bea 
C^a_{e[b}C^e_{cd]}&=&e^i_{[b}\partial_iC^a_{cd]}~,\label{nc1}
\\
2e^j_{[b}\partial_je^i_{c]}&=&C^a_{bc}e^i_a~,\label{nc2}
\eea
where here and in the following symmetrizations and antisymmetrizations of indices are taken with the 
corresponding weight. 
In the special case that the coefficients $C^a_{bc}$ are constant,
the right hand side of the first condition vanishes and the condition simply becomes the usual Jacobi identity for 
 a Lie algebra $\mf g$. Moreover, we then can define the $\mf g$-valued Maurer-Cartan form 
 $e=e^ae_a\in \Omega^1(N,\mf g)$ which 
 satisfies the famous Maurer-Cartan equation 
 \be 
 \dd e+\sfrac 12 [e,e]=0~.
 \ee
  However, 
when $C^a_{bc}$ are not constant, but rather $\varphi$-dependent functions, one may address 
the problem in the following way, cf. \cite{crainic}.
 First one considers a vector bundle $L$ of rank $r$ over $U$.
$L$ is equipped with a Lie bracket $[\cdot,\cdot]_L$ and $C^a_{bc}$ become structure functions for 
this bracket, namely 
\be 
[e_a,e_b]_{L}=C^c_{ab}(\varphi)e_c~,\label{bracket}
\ee
for a basis of local sections $e_a\in\G(L)$.
Furthermore, a smooth map $\rho: L \to TU$ is introduced, mapping sections of $L$ to vector fields. 
The bracket on $L$ satisfies the Leibniz identity 
$[e,fe']_L=f[e,e']_L+\rho(e)f\, e'$, for $e,e'\in\G(L)$ and $f\in C^{\infty}(U)$. 
 Then Eq. \eqref{nc2} 
is simply the statement that this map is a homomorphism:
\be 
\eqref{nc2} \quad \Leftrightarrow \quad \rho([e_a,e_b]_L)=[\rho(e_a),\rho(e_b)]~.
\ee
Moreover, Eq. \eqref{nc1} simply states that:
\be 
\eqref{nc1} \quad \Leftrightarrow \quad \text{Jac}(a,b,c):=[e_a,[e_b,e_c]_L]_L ~ + ~ \text{(cyclic permutations)}=0~;
\ee
in other words it is the Jacobi identity for the Lie bracket on $L$. The above triple $(L,[\cdot,\cdot]_{L},\rho)$ is an example 
of a Lie algebroid. More generally, a Lie algebroid over a manifold $M$ consists of
 a vector bundle $L$ over $M$, equipped with a Leibniz-Lie bracket on its sections $\G(L)$ and a bundle map $\rho: L\to TM$. 
Lie algebras are included in the definition; they are the corresponding structures in the 
case that the base manifold is just a point. 

%---2nd fold 

Returning to our discussion of the $\sigma$-model \eqref{string}, let us take a step back and recall that given a set of vector fields $\rho_a=\rho_a^i(X)\partial_i$, the action \eqref{string} has a global symmetry 
of the form
\be 
\d_{\epsilon}X^i=\rho^i_a(X)\epsilon^a~\label{sym}
\ee
for rigid transformation parameters $\epsilon^a$ only under the following three invariance conditions: 
\bea 
{\cal L}_{\rho_a}g=0~, \label{isog} \quad 
{\cal L}_{\rho_a}B=\dd\hat\theta_a~,\quad
{\cal L}_{\rho_a}\Phi=0~,
\eea
for some  1-form $\hat\theta_a$. Then $\rho_a$ are Killing vector fields generating isometries for the metric $g$.
They can be non-Abelian and satisfy an algebra 
$$
[\rho_a,\rho_b]=C_{ab}^c\,\rho_c~.
$$ 
The symmetry is gauged through $\mf{g}$-valued  1-forms $A=(A^a)$. The gauge parameters $\epsilon^a$ now depend on the world sheet coordinates $(\s^{\a})$ and the gauge fields transform according to 
\be 
\d_{\epsilon}A^a=\dd\epsilon^a+C^a_{bc}A^b\epsilon^c~,\label{AgaugeLie}
\ee 
which is the standard infinitesimal gauge transformation for non-Abelian gauge fields. Clearly,  
the above invariance conditions should also hold at the level of the gauged model, and in addition $\hat\theta_a$ 
has to be zero. The latter fact is related to the validity of minimal coupling and will be discussed below. 
 
However, \eqref{isog} are too strong and restrictive. We address this issue by following 
a more unorthodox route and asking the question
\begin{itemize}
\item Under which invariance conditions does the action functional \eqref{string} have a local symmetry 
$\d_{\epsilon}X^i=\rho_a^i(X)\epsilon^a(\sigma)$ at leading order in $\alpha'$?
\end{itemize}
As already mentioned, the main difference is that we do not require a global symmetry to start with, thus 
allowing for more general background fields. To answer this question we are going to consider 
gauge fields taking values in a Lie algebroid 
\be 
L \overset{\rho}\to TM~.
\ee
The analysis reveals that the invariance conditions \eqref{isog} are 
replaced by the milder ones  \cite{cdjs} 
\bea 
{\cal L}_{\rho_a}g &= &\omega_a^b\vee\iota_{\rho_b}g-\phi_a^b\vee \iota_{\rho_b}B ~,\label{ic1}
\\
{\cal L}_{\rho_a}B &= &\omega_a^b\w\iota_{\rho_b}B\pm \phi_a^b\w\iota_{\rho_b}g~,\label{ic2}
\eea
where $\vee$ is defined as the symmetric product $\a\vee\beta=\a\otimes\beta+\beta\otimes\a$ and $\wedge$ as the 
antisymmetric one $\a\w\beta=\a\otimes\beta-\beta\otimes\a$, while $\omega_a^b=\omega_{ai}^b\dd x^i$ and 
$\phi_a^b=\phi_{ai}^b\dd x^i$ are 1-forms on $M$, where $(x^i)$ is a local coordinate system on it.
In a geometric interpretation of these relations we find that $\omega_a^b$ and $\phi_a^b$ in fact define two connections
$\nabla^{\pm}$ on $L$. 
This is our second generalization.

%---3rd fold

Next let us return to the issue of minimal coupling. As mentioned above, when $\hat\theta_a$ in 
\eqref{isog} is not zero, minimal coupling is not enough. This is certainly not a new observation. 
It is related to the presence of Wess-Zumino (WZ) terms and has been clarified in many works 
starting from \cite{hs1,hs2,fs} and more recently in \cite{Plauschinn1,Plauschinn2}. One of the main points is that when 
there is a WZ term one has to add terms in the gauged action which contain the gauge fields 
outside the covariant derivatives that are formed through minimal coupling. Then the gauging 
is obstructed by additional constraints on top of the previously mentioned invariance 
conditions. 
Revisiting this issue in the light 
of the two generalizations we already mentioned leads us again to a set of milder conditions 
and constraints for WZ terms. Furthermore, it 
turns out that the target space of the gauged theory 
is lifted to the generalized tangent bundle $TM\oplus T^{\ast}M$ \cite{cdjs}.\footnote{Cf. \cite{Plauschinn1,kss} 
for the analogous result in a less general context.}

%---Gauge symmetries of DSMs

A very interesting question arising from the above results is whether and when the new and milder invariance conditions for the gauging of the $\sigma$-model can be solved. We 
will show that when $L$ is a Dirac structure, i.e. a particular type of Lie algebroid 
associated to the standard Courant algebroid on $TM\oplus T^{\ast}M$ twisted by a closed 3-form $H$, then the two Dirac structure connections 
solving the invariance conditions can be determined. Indeed we are going to provide general closed expressions for them. This is the central result of this paper.
Additionally, it will allow us to find the form of the gauge symmetries for Dirac $\sigma$-models, which are very general two-dimensional topological field theories interpolating 
between WZW models and Poisson $\sigma$-models and they were introduced in \cite{dsm1,dsm2}. Our analysis will be complemented 
with an explicit example 
where we will apply our findings to the WZ Poisson $\sigma$-model with a kinetic term.

\section{General gauge invariant 2D $\sigma$-models}
\label{Sec.2}

In this section we present some of the material from \cite{cdjs} in a self-contained manner, since it is needed as a basis for the subsequent sections. This applies in particular to sections \ref{Sec.2.1} and \ref{nonmin}. Sec.~\ref{Sec2.2}, on the other hand, provides some relation of the obtained formulas to generalized geometry; for the complete picture of this we refer to \cite{ks3}.

\subsection{Gauge theory and invariance conditions for minimally-coupled gauge fields}
\label{Sec.2.1}

As explained in the introduction, our starting point is 
the bosonic sector of closed string theory at leading order in the string slope parameter $\a'$. 
This is described by the following $\sigma$-model action:
  \bea\label{s0}
 S_0[X]&=&\int_{\Sigma}\sfrac 12 g_{ij}(X)\dd X^i\wedge \ast \dd X^j+\int_{\Sigma} \sfrac 12 B_{ij}(X)\dd X^i\wedge\dd X^j~,
 \eea
 where $\Sigma$ is the 2D world sheet, $X=(X^i):\Sigma\to M, i=1,\cdots,n=\text{dim}M,$ is the map from the 
 world sheet to the target space $M$ and $\ast$ denotes the Hodge duality operator on the world sheet. 
 Here and in the following we work in units where $4\pi\a'=1$. The background fields
 $g=g_{ij}\dd x^i\otimes \dd x^j=\sfrac 12 g_{ij}\dd x^i\vee\dd x^j$ and $B=\sfrac 12 B_{ij}\dd x^i\w\dd x^j$ are the metric and 
 the antisymmetric Kalb-Ramond 2-form on $M$, where $(x^i)$ are local coordinates on it. Then $g(X)=X^{\ast}g$ and 
 $B(X)=X^{\ast}B$ are the pull-back metric and $B$-field by means of the map $X$. 
 In what follows we ignore the dilaton coupling, which enters 
 the action at linear order in $\alpha'$.
 
 Now we would like to find a gauged extension of the action \eqref{s0} for as general $X$-dependent background fields $g(X)$ and 
 $B(X)$ as possible. To this end we introduce a set of world sheet 1-forms $A$ taking values in a Lie algebroid $L\overset{\rho}\to TM$. 
 Let us denote local sections of $L$ by $e_a$, $a=1,\dots,r$, $r$ being the rank of the vector bundle $L$. These sections are mapped 
 via the smooth map $\rho$ to a set of vector fields $\rho_a$:
 \bea 
 L&\overset{\rho}\to& TM \nn\\
 e_a &\mapsto& \rho(e_a):=\rho_a=\rho_a^i(X)\partial_i~.\label{rho}
 \eea
 Recalling that $L$ is equipped with a bracket such that 
 \be \label{alg}
 [e_a,e_b]_L=C_{ab}^c(X)e_c~,
 \ee
 and the map $\rho$ is a homomorphism, we deduce that the vector fields $\rho_a$ satisfy 
 \be 
 [\rho_a,\rho_b]=C_{ab}^c(X)\rho_c~,
 \ee
 the bracket being the standard Lie bracket of vector fields. As for the gauge fields $A=A^ae_a \in\G(L)$, these are mapped to $\rho(A)\in\G(TM)$.
 In this subsection we examine the case when these gauge fields are introduced in the theory through minimal coupling. 
 In other words we promote the world sheet differential to a covariant one as 
 \be \label{cD}
 \dD X^i=\dd X^i-\rho^i_a(X)A^a~,
 \ee
 or in index-free notation $\dD X=\dd X-\rho(A)$. 
 The map $\rho$ need not be invertible and in general it is not. For example, if we choose to gauge 
 the theory only along $d\leq n=\text{dim}M$ directions of the target, then in adapted coordinates 
 where 
 the set of spacetime indices $\{i\}$  splits into $\{\mu,m\}$, 
 for $\m=1,\ldots,n-d$ and $m=n-d+1,\ldots, n$, we would have $\rho_a^\mu=0$ and consequently $\dD X^\mu=\dd X^\mu$. 
 In any case, minimal coupling then means that the gauge fields appear in the gauged theory only through the 
 covariant derivative $\dD X^i$.

According to the above, the action we consider is
 \bea\label{sgauged}
 S_{\text{m.c.}}[X,A]&=&\int_{\Sigma}\sfrac 12 g_{ij}(X)\dD X^i\wedge \ast \dD X^j
 +\int_{\Sigma} \sfrac 12 B_{ij}(X)\dD X^i\wedge \dD X^j~.
 \eea
Then the question is under which conditions does this action have a gauge symmetry generated by the vector fields $\rho_a$, namely
\be\label{dX}
\d_{\epsilon}X^i=\rho^i_a(X)\epsilon^a(\s)~,
\ee
where $\epsilon^a\in C^{\infty}(\S)$ is the gauge transformation parameter. This crucially depends on how the gauge fields $A^a$ transform. 
Let us consider a general Ansatz for their gauge transformation:
\be \label{dA}
\d_{\epsilon} A^a=\dd \epsilon^a+C^a_{bc}(X)A^b\epsilon^c+\D A^a~,
\ee
where for the moment the additional term $\D A^a$, which of course has to be a world sheet 1-form, is not specified.
Now we can determine how the covariant derivative transforms under these gauge transformations. It is found that 
\be 
\d_{\epsilon}\dD X^i=\epsilon^a\partial_j\rho^i_a\dD X^j-\rho^i_a\D A^a~.
\ee
Requiring a covariant transformation rule for the covariant derivative motivates us to refine the Ansatz by writing 
\be \label{DA}
\D A^a=\omega^a_{bi}(X)\epsilon^b\dD X^i+\phi^a_{bi}(X)\epsilon^b\ast\dD X^i~.
\ee
First of all, let us mention that the addition of the second term is possible only in two dimensions, where the Hodge dual of an 1-form is 
also an 1-form. Second, the coefficients $\omega^a_{bi}$ and $\phi^a_{bi}$ are still undetermined functions defined locally 
on $M$ and their geometric interpretation will be clarified in due course. 

Next let us examine the behaviour of the action \eqref{sgauged} under the above gauge transformations. A direct computation leads to 
the result
\bea 
\d_{\epsilon}S_{\text{m.c.}}&=&\int_{\S}\epsilon^a\left(\sfrac 12 \left({\cal L}_{\rho_a}g\right)_{ij}\dD X^i\w\ast\dD X^j+ 
\sfrac 12 \left({\cal L}_{\rho_{a}}B\right)_{ij}\dD X^i\w\dD X^j\right)-\nn\\
&& -\int_{\S}\left(g_{ij}\rho^i_a\D A^a\w\ast\dD X^j+B_{ij}\rho^i_a\D A^a\w\dD X^j\right)~.
\eea
It is directly observed that in case $\D A^a=0$, namely when $A^a$ transforms as a standard non-Abelian gauge field, 
the results reviewed in the introduction are immediately recovered. However, now a new possibility is revealed. 
Due to \eqref{DA} the terms in the second line of the transformed action can compensate for the ones in the first line. 
Indeed, defining the 1-forms $\omega^a_b=\omega^a_{bi}\dd x^i$ and $\phi^a_b=\phi^a_{bi}\dd x^i$, the action functional is gauge invariant 
if and only if the following two conditions hold:
\begin{empheq}[box=\fbox]{align}
\label{cg1} {\cal L}_{{\rho}_a}g&=\o^b_a\vee\iota_{{\rho}_b}g-\phi^b_a\vee \iota_{\rho_b}B~,
\\
\label{cB1} {\cal L}_{{\rho}_a}B &= \o^b_a\w \iota_{\rho_b}B\pm \phi^b_a\w \iota_{{\rho}_b}g~,
\end{empheq}
where the $\pm$ sign is due to $\ast^2=\mp 1$ and we recall that $\vee$ denotes the symmetrized tensor product; thus for example 
$(\omega^a_b\vee \iota_{\rho_b}g)_{ij}=\omega^b_{ai}\rho^k_bg_{kj}+\omega^b_{aj}\rho^k_bg_{ki}$. 
Clearly \eqref{cg1} is a symmetric equation and \eqref{cB1} an antisymmetric one.

We observe that the  action \eqref{sgauged} can be gauge invariant under the transformation \eqref{dX} without the vector fields 
$\rho_a$ being Killing, i.e. without them generating isometries. This is true provided the gauge fields transform according to 
\eqref{dA}. Thus the resulting invariance conditions are milder than in the case of conventional gauging. Furthermore, 
it is worth noting that due to the additional term in \eqref{dA}, which is proportional to the Hodge dual covariant differential, 
the invariance conditions mix the metric and the Kalb-Ramond field. This is indicative of an interpretation in terms of generalized geometry, 
as we discuss immediately below. 

\subsection{Geometric interpretation and generalized Riemannian metric}
\label{Sec2.2}

Having found the extended invariance conditions \eqref{cg1} and \eqref{cB1} that guarantee gauge invariance of the action, we would 
like to explain the geometric role of the coefficients $\omega^a_{bi}$ and $\phi^a_{bi}$. 
In order to understand their geometric interpretation let us examine what happens under a change of basis in the vector bundle $L$. 
First note that when we change the basis $e_a\to \Lambda(X)^b_a e_b$ the gauge field transforms as 
$A^a\to (\Lambda^{-1}(X))^a_bA^b$ so that $A=A^ae_a$ remains invariant. 
Next, from the transformation of \eqref{alg} we determine the transformation properties of the structure functions $C^a_{bc}(X)$:
\bea\label{Ct}
C^a_{bc}\to (\Lambda^{-1})^a_d\Lambda^e_b\Lambda^f_c C^d_{ef}+2 (\Lambda^{-1})^a_d\Lambda^e_{[b}\rho^i_{\underline e}\partial_i
\Lambda^d_{c]}~,
\eea
where the underlined index does not participate in the antisymmetrization. 
We notice that the structure functions do not transform as tensors under the change of basis, as expected (cf. \cite{ks2}).

Demanding that the form of the gauge transformation \eqref{dA} does not change under the change of local basis we obtain
the transformation properties of  the coefficients $\o^a_{bi}$ and $\phi^a_{bi}$:
\bea\label{transo}
\o^a_{bi}&\to &(\Lambda^{-1})^a_c\o^c_{di}\Lambda^d_b-\Lambda^c_b\partial_i(\Lambda^{-1})^a_c~,\\
\phi^a_{bi}&\to &(\Lambda^{-1})^a_c\phi^c_{di}\Lambda^d_b~. \eea
It is directly observed that $\phi^a_b$ transforms as a tensor, however $\omega^a_b$ does not; it transforms instead 
as a connection, as noticed before in \cite{ks2}. This means that $\omega^a_{bi}$ are the coefficients of a connection 1-form on $L$: 
\be 
\nabla^{\omega} e_a=\omega_a^b\otimes e_b~,
\ee
whilst $\phi^a_b$ are the components of an endomorphism-valued 1-form $\phi\in\G(T^{\ast}M\otimes E^{\ast}\otimes E)$. 
Since the difference of any two connections on a vector bundle is such an endomorphism-valued 1-form on the manifold 
$M$, this essentially means that we have two connections on $L$, $\nabla=\nabla^{\o}$ and $\widetilde\nabla=\nabla^{\o}+\phi$. 
In fact it is more convenient to introduce the connections 
\be 
\nabla^{\pm}=\nabla^{\o}\pm\phi~,
\ee 
for the reason that is explained immediately below.

In order to gain a better geometrical understanding of the mixing of $g$ and $B$ in the invariance 
conditions \eqref{cg1} and \eqref{cB1}, it is useful to consider the two maps 
\be 
E^{\pm}:=B\pm g : TM \to T^{\ast}M~,
\ee
defined via the interior product. Additionally, we define the combinations 
\be 
(\Omega^{\pm})^a_b:=(\omega\pm\phi)^a_b~.
\ee 
Then the two invariance conditions for the Lorentzian signature of the world sheet metric 
may be expressed as 
\bea\label{add}
{\cal L}_{{\rho}_a}E^{\pm}=(\Omega^{\mp})_a^b\otimes \iota_{\rho_b}E^{\pm}-\iota_{\rho_b}E^{\mp}\otimes (\Omega^{\pm})^b_a~.
\eea
This expression highlights the role of the two connections defined above. Moreover, let us recall from \cite{gg1,gg2} the role of the maps 
$E^{\pm}$. In generalized complex geometry one deals with structures on the generalized tangent bundle $TM\oplus T^{\ast}M$, whose 
structure group is $O(n,n)$. A generalized Riemannian metric is a reduction of the structure group to 
$O(n)\times O(n)$ or equivalently a choice of a $n$-dimensional subbundle $C_+$ which is positive definite with 
respect to the natural inner product on $TM\oplus T^{\ast}M$ and whose negative-definite 
orthogonal complement is $C_-$. Then (cf. Proposition 
6.6 in \cite{gg1}) the subbundles $C_{\pm}$ are identified with the graphs of the maps $E^{\pm}$. Using these data one can define a 
generalized metric ${\cal H}:TM\oplus T^{\ast}M\to TM\oplus T^{\ast}M$ on the generalized tangent bundle, which in terms of $g$ and $B$ 
takes the form
\be 
{\cal H}=\begin{pmatrix}
        -g^{-1}B  & g^{-1} \\ g-Bg^{-1}B& Bg^{-1}
         \end{pmatrix}~,
\ee
where $g^{-1}$ is the inverse metric. We refer to \cite{gg2} for more details, since we are not going to use this metric any more in this 
paper. However, it is important to keep in mind the appearance of the generalized tangent bundle since it will play an interesting role 
at the end of this section, after we discuss the inclusion of WZ terms in the theory.

\subsection{Inclusion of WZ term and non-minimally coupled gauge fields}\label{nonmin}

Let us now go one step further and include a WZ term in the $\s$-model action. Therefore the original ungauged action now 
is
\be 
S_{0,\text{WZ}}[X]=\int_{\S}\sfrac 12 g_{ij}(X)\dd X^i\w\ast \dd X^j+\int_{\hat{\S}}\sfrac 16 H_{ijk}(X)\dd X^i\w\dd X^j\w\dd X^k~,
\ee 
where $\hat{\S}$ is an open membrane world volume whose boundary is the world sheet $\S$, $\partial\hat \S=\S$. 
As before $H(X)=X^{\ast}H$, where $H=\sfrac 16 H_{ijk}\dd x^i\w\dd x^j\w\dd x^k$ is a closed 3-form on $M$. This 3-form need not be 
exact at the global level, thus $H=\dd B$ only locally.
Before proceeding with our analysis, let us recall some well-known facts from the ordinary gauging of this action. First of all, 
in the presence of WZ terms minimal coupling of the gauge fields is not sufficient to guarantee a gauge invariant action. 
This means that terms with ``bare'' gauge fields, i.e. outside of covariant derivatives, have to be added to the 
topological sector of the action. On the other hand this is not necessary for the kinetic sector, where gauge fields can still enter through minimal 
coupling.{\footnote{Of course it is possible to also consider non-minimally coupled kinetic terms but we do not see sufficient
motivation for this 
in the context of the present paper. The investigation of the most general case will be presented in \cite{cdjs}.}} Second, already in the isometric case the invariance conditions on the background fields do not 
automatically render the action gauge invariant. There are additional constraints that have to be met. This will be true also in 
our analysis here.

 In order to address these issues in the more general context that we employ in this paper, 
 we start from a candidate gauge action of the form
\begin{equation} \label{SWZ}
S_{\text{WZ}}[X,A]=\int_{\Sigma}\sfrac 12 g_{ij} (X)\dD X^i \w \ast \dD X^j + \int_{\hat\Sigma} H(X)+
\int_{\Sigma} A^a \wedge \theta_a +\int_{\Sigma}  \sfrac{1}{2}\gamma_{ab} A^a \wedge A^b~, 
\end{equation}
where $\theta_a=\theta_{ai}(X)\dd X^i$ are 1-forms and $\gamma_{ab}$ functions on the target space $M$, 
both pulled back by $X \colon \Sigma \to M$.
The covariant derivative $\dD X^i$ is again defined as in \eqref{cD}, with gauge fields taking values in the vector bundle $L$. 
Once more we ask under which conditions this action is invariant under the gauge transformations \eqref{dX} and \eqref{dA}, which 
we repeat here for completeness:
\bea 
\label{dXre}
\d_{\epsilon}X^i&=&\rho^i_a(X)\epsilon^a(\s)~,\\
\label{dAre}
\d_{\epsilon} A^a&=&\dd \epsilon^a+C^a_{bc}(X)A^b\epsilon^c+\omega^a_{bi}(X)\epsilon^b\dD X^i+\phi^a_{bi}(X)\epsilon^b\ast\dD X^i~.
\eea
Transforming the action \eqref{SWZ} under these gauge transformations we find 
\bea\label{var S}
\delta_\epsilon S_{\text{WZ}} = \int_{\S} \epsilon^a \hspace{-0.5cm}&& \left\{
\sfrac 12\left({\cal L}_{{\rho}_{a}}g-\omega^b_a\vee\iota_{\rho_b}g-\phi^b_a\vee\theta_b\right)_{ij}\dD X^i\w\ast\dD X^j\right.+\nn\\
 &&+\sfrac 12
\left( \iota_{\rho_a}H-\dd\theta_a+\omega^b_a\w\theta_b\mp\phi^b_a\w\iota_{\rho_b}g\right)_{ij}\dD X^i\w \dD X^j+\nn\\
&&- \left( {\cal L}_{\rho_a} \theta_b - C^c_{ab} \theta_c +\iota_{\rho_b} (\iota_{\rho_a}H-\dd\theta_a )+(\gamma_{bd}-\iota_{\rho_b}\theta_d 
)\omega^d_a\right)_i \dd X^i\w A^b+\nn\\
&&\left.+\left(\sfrac 12{\cal L}_{\rho_a}\gamma_{bc}+C^d_{ab}\gamma_{cd}-\sfrac 12\iota_{\rho_c}\iota_{\rho_b}(\iota_{\rho_a} H
-\dd\theta_a)+(\gamma_{bd}-\iota_{\rho_b}\theta_d)\iota_{\rho_c}\o^d_a\right)A^b\w A^c\right\}+\nn\\
+\int_{\S} \hspace{-0.5cm}&&(\gamma_{ab}-\iota_{\rho_a}\theta_b)(\dd\epsilon^a\w A^b+\epsilon^c\phi^b_{ci}\dD X^i\w\ast A^a)~.
\eea
Gauge invariance is established when all five lines in the above result vanish. However, the third and the fifth line 
together imply the fourth one, while the fifth one itself states that $\gamma_{ab}=\iota_{\rho_a}\theta_b$. 
Thus we conclude that the functional \eqref{SWZ} is invariant with respect to gauge 
transformations of the form \eqref{dXre} and \eqref{dAre} if and only if the following four equations hold true:  
\begin{empheq}[box=\fbox]{align}
\label{cg} {\cal L}_{{\rho}_a}g&=\o^b_a\vee\iota_{{\rho}_b}g+\phi^b_a\vee \theta_b~,
\\
\label{cH} \iota_{\rho_a} H &=\dd \theta_a - \o^b_a\w\theta_b \pm \phi^b_a\w \iota_{{\rho}_b}g~,
\end{empheq}
supplemented by the constraints
\bea \label{cc1}
&&\iota_{\rho_a}\theta_b +\iota_{\rho_b}\theta_a =0~,\\
&&{\cal L}_{\rho_a}\theta_b=C^d_{ab}\theta_d-\iota_{\rho_b}(\iota_{\rho_a}H-\dd\theta_a)~.\label{cc2}
\eea
The last constraint can be rewritten in the form of a double contraction on $H$:
\be \label{cc2b}
\iota_{\rho_b}\iota_{\rho_a}H=C^d_{ab}\theta_d+\dd\iota_{\rho_{[a}}\theta_{b]} -2{\cal L}_{\rho_{[a}}\theta_{b]}~,
\ee
an expression which will acquire a natural geometric interpretation below.

When both $\omega^a_b$ and $\phi^a_b$ vanish, the known results for the non-Abelian but isometric case 
are recovered \cite{hs1,hs2,Alvarez1,Alvarez2,Plauschinn1,Plauschinn2}. However we observe that 
in general the conditions \eqref{cg} 
and \eqref{cH} as well as the additional constraint \eqref{cc2} are milder than in the standard case. (The constraint 
 \eqref{cc1} remains the same though.)
 
 \paragraph{Addendum on the geometric interpretation.} The geometric interpretation of these results is similar to the minimally-coupled case; there are two connections 
 $\nabla^{\pm}$ on $L$ that control the gauging. However, in the present case the role of the generalized tangent bundle 
 is further clarified. Indeed, in the spirit of \eqref{rho} for the map $\rho$, the 1-form $\theta_a$ may be 
 associated to the following map $\theta$ from the vector bundle $L$ to the cotangent bundle of $M$:
 \bea 
 L&\overset{\theta}\to& T^{\ast}M \nn\\
 e_a &\mapsto& \theta(e_a):=\theta_a=\theta_{ai}\dd x^i~.\label{theta}
 \eea
 Combining the two maps we obtain
  \bea 
 L&\overset{\rho\oplus\theta}\to& TM\oplus T^{\ast}M \nn\\
 e_a &\mapsto& (\rho\oplus\theta)(e_a):=\rho_a+\theta_a=\rho^i_a\partial_i+\theta_{ai}\dd x^i~.\label{rhotheta}
 \eea
 This is reminiscent of a Courant algebroid structure on the generalized tangent bundle, in particular the $H$-twisted 
 standard Courant algebroid. Local sections of $TM\oplus T^{\ast}M$ are generalized vectors $\xi_a=\rho_a+\theta_a$ 
 and the corresponding bracket 
 of sections is the Courant bracket
 \be \label{courant}
 [\xi_a, \xi_b]=[\rho_a,\rho_b]+{\cal L}_{\rho_a}\theta_b-{\cal L}_{\rho_b}\theta_a-\sfrac 12 \dd \left(\iota_{\rho_a}\theta_b-
 \iota_{\rho_b}\theta_a\right)-\iota_{\rho_a}\iota_{\rho_b}H~.
 \ee
 Then the constraint \eqref{cc2} is interpreted as closure of this $H$-twisted Courant
bracket for the generalized vectors $\xi_a$. This is more easily verified using the 
equivalent expression \eqref{cc2b}. Moreover, the constraint \eqref{cc1} is equivalent to the statement that 
the generalized vectors $\xi_a$ have a vanishing non-degenerate symmetric bilinear form 
\be 
\langle \xi_a,\xi_b\rangle= \iota_{\rho_a}\theta_b+\iota_{\rho_b}\theta_a~. 
\ee
In other words the two constraints require that $\xi_a$ are sections of an involutive and isotropic subbundle of
$TM\oplus T^{\ast}M$. (This was also noticed in a similar context in \cite{kss} and also in \cite{Plauschinn2}, 
where in the presence of additional scalar fields it was possible to relax the isotropy condition.)
Such subbundles are called (small) Dirac structures and will be studied in more detail in the upcoming section.

\section{Connections and gauge symmetries for Dirac structures}
\label{Sect.3}

\subsection{Dirac structures and Dirac $\s$-models}

 In the previous section we found that the extended invariance conditions \eqref{cg} and \eqref{cH} on the background fields, 
 which are necessary for gauge invariance, are controlled by the coefficients $\omega^a_{bi}$ and $\phi^a_{bi}$ which give rise to 
 two connections on the vector bundle $L$. A natural question is whether these coefficients can be determined explicitly, or in other words, whether we can 
 really provide the corresponding connections. Furthermore, at the end of the last section we saw that the additional constraints 
 obstructing the gauging of the $\sigma$-model are associated to a particular class of subbundles in the $H$-twisted 
 Courant algebroid. 
 
 In order to study the construction of the two connections, we now focus on Dirac structures. As indicated already, these are 
 maximal subbundles $D\subset TM\oplus T^{\ast}M$ of the generalized tangent bundle with the following two properties:
 \bea 
 [\G(D),\G(D)]&\subset& \G(D)~,\\
 \langle\G(D),\G(D)\rangle&=&0~,
 \eea
 namely they are involutive with respect to the $H$-twisted Courant bracket and isotropic with respect to the symmetric bilinear form on the Courant algebroid. 
 They were introduced in \cite{dirac} as generalization of symplectic and Poisson structures.
 Maximality in this case means that their rank is half of the rank of $TM\oplus T^{\ast}M$. Moreover, since maximality is not 
 indicated by any of the found constraints, one may define a small Dirac structure as being any subbundle satisfying the above two conditions. 
 This definition first appeared in \cite{kss}.
 
 At the availability of a metric $g$ on $M$, which is the case here, yet another parametrization of Dirac structures is possible, 
 as explained in \cite{dsm1,dsm2}. Instead of referring to the generalized vectors $\xi_a=\rho_a+\theta_a$, one uses the 
 Riemannian metric $g$ to identify $TM$ with $T^{\ast}M$ and introduces an orthogonal operator
 ${\cal O}={\cal O}^i_{\ j}\partial_i\otimes \dd X^j\in\G(\text{End}(TM))$ whose 
graph is the Dirac structure $D$. The role of this operator is that 
for any element, say $v\oplus \eta$ with $v$ a vector and $\eta$ an 1-form, of the Dirac structure there exists a unique section $\mf a$ of $TM$ such that 
\be \label{Aa}
v\oplus \eta=(\text{id}-{\cal O})\mf a\oplus \left(\left(\text{id}+{\cal O}\right)\mf a\right)^{\ast}~,
\ee 
where the notation $\upsilon^{\ast}$ and $\eta^{\ast}$ for a vector field $\upsilon$ or a 1-form $\eta$ 
denotes the action of the metric or inverse metric that results in an 
1-form or a vector field respectively, explicitly
\be 
\upsilon^{\ast}=(\upsilon^i\partial_i)^{\ast}=\upsilon^ig_{ij}\dd x^j~,\quad \eta^{\ast}=(\eta_i\dd x^i)^{\ast}=\eta_ig^{ij}\partial_j~.
\ee

Given a Dirac structure $D$ one may uniquely construct a two-dimensional $\sigma$-model called 
Dirac $\s$-model (DSM) \cite{dsm1}, which was introduced as a simultaneous generalization of  
the Poisson sigma model \cite{psm1,psm2} and the G/G WZW model. Its action functional 
is 
\be\label{SDSM} S_{\text{DSM}}[X,v\oplus \eta]=\int_{\S}\sfrac 12 g_{ij}(X)\dD X^i\w\ast \dD X^j
	+\int_{\S}\left(\eta_i\w\dd X^i-\sfrac 12 \eta_i\w v^i\right)+\int_{\hat\S}H~,
\ee
where $v\oplus \eta\in\Omega^1(\S,X^{\ast}D)$ and here we defined the world sheet covariant differential as
$\dD X^i=\dd X^i-v^i$. In terms of the alternative parametrization introduced above, i.e., here with $ \mf a\in \Omega^1(\S,X^{\ast}TM)$, an equivalent form of this action, already presented in Ref. \cite{dsm1}, is
\bea
S_{\text{DSM}}[X,\mf a] &=& 
\int_\Sigma \sfrac{1}{2}g_{ij} \left(\dd X^i-\left(\text{id}-{\cal O}\right)^i_{\ k}\mf a^k\right)
  \w \ast\left(\dd X^j-\left(\text{id}-{\cal O}\right)^j_{\ l}\mf a^l\right) + \nn\\ 
&& +  \int_\Sigma \left(\left(g_{ij}+{\cal O}_{ij}\right) \mf a^j \w \dd X^i+{\cal O}_{ij}\mf a^i\w \mf a^j\right)+ \int_{\hat\S} H  \: , 
\eea
with indices lowered by means of the metric $g$. Even though there is a kinetic sector, this model turns 
out to be a topological field theory \cite{dsm1}. A non-topological field theory can be constructed by assuming instead 
a small (non-maximal) Dirac structure \cite{kss}. 

An interesting observation, already suggested in \cite{kss}, albeit in a less general context, is that upon the relations
\be 
v=\rho(A)\Rightarrow v^i=\rho^i_a(X)A^a \quad \text{and} \quad \eta=\theta(A)\Rightarrow \eta_i=\theta_{ai}(X)A^a~,
\ee 
the action functional \eqref{SWZ} takes the form \eqref{SDSM}. This relation among the two action 
functionals is very useful in constructing the desired connections and determining the general form of the 
gauge symmetries for the DSM \eqref{SDSM}. The latter were suggested without proof in 
the original publication \cite{dsm1}, but here we will prove them at the end of this section.

\subsection{Constructing the connections $\nabla^{\pm}$ on Dirac structures}

Our interest now is to determine explicitly the two connections $\nabla^{\pm}$ in the case that both $L=D$ and its 
image $\widetilde D$ under the map $\rho\oplus\theta$ in \eqref{rhotheta} are Dirac structures in 
the $H$-twisted standard Courant algebroid. 

First, it is useful to recall a lemma proven in \cite{dsm1} (Lemma 1), from which it follows that the operator 
\be 
(\text{id}-{\cal O})+b(\text{id}+{\cal O})~,
\ee
where $b$ is positive or negative symmetric operator,
is invertible. 
Then let us consider the sections $v\oplus \eta\in \Omega^1(\S,X^{\ast}D)$ on the Dirac structure $D$
and $\widetilde v\oplus \widetilde \eta\in \Omega^1(\S,X^{\ast}\widetilde D)$ on the Dirac structure $\widetilde D$. 
Recalling the parametrization in terms of the orthogonal operator, we parametrize $D$ by ${\cal O}$
and $\widetilde D$ by $\widetilde{\cal O}$. These considerations allow us to write
\be 
v=(\text{id}-{\cal O})\mf a~, \quad \eta=\left(\left(\text{id}+{\cal O}\right)\mf a\right)^{\ast}~, 
\ee 
and 
\be 
\widetilde v=(\text{id}-\widetilde {\cal O})\mf a~, \quad \widetilde \eta=\left((\text{id}+\widetilde {\cal O})\mf a\right)^{\ast}~,
\ee 
for some $ \mf a\in \Omega^1(\S,X^{\ast}TM)$.
Moreover by assumption it holds that
\bea 
\widetilde v=\rho(v\oplus \eta)~, \quad \text{and} \quad \widetilde \eta=\theta(v\oplus \eta)~.
\eea
These relations directly translate into
\bea 
\text{id}-\widetilde {\cal O}&=&\rho\left((\text{id}-{\cal O})+(\text{id}+{\cal O})^{\ast}\right)~,\\
\text{id}+\widetilde {\cal O}&=&\left(\theta\left((\text{id}-{\cal O})+(\text{id}+{\cal O})^{\ast}\right)\right)^{\ast}~,
\eea
which in turn yield
\bea 
\sfrac 12 (\theta^{\ast}+\rho)\left((\text{id}-{\cal O})+(\text{id}+{\cal O})^{\ast}\right)&=&\text{id}~,
\\
\sfrac 12 (\theta^{\ast}-\rho)\left((\text{id}-{\cal O})+(\text{id}+{\cal O})^{\ast}\right)&=&\widetilde {\cal O}~.
\eea
Now the right hand side of both equations is an invertible operator. 
Thus we conclude that the operators 
\be 
\theta^{\ast}\pm\rho: D \to TM~,
\ee 
are invertible too, with inverses denoted as $(\theta^{\ast}\pm\rho)^{-1}:TM\to D$. We note in passing that certainly none of the maps $\rho$ and $\theta^{\ast}$ are required 
to be invertible separately. 

Now let us state and prove the main result of this section. For the vector bundles $D$ and $\widetilde D$ as above, the invariance 
conditions \eqref{cg} and \eqref{cH} are solved by the coefficients
\bea 
\omega^{a}_{bi}&=&\G^a_{bi} - \phi^a_{bi}+T^a_{bi}~,
\label{omegagen} \\
\phi^a_{bi}&=&[(\theta^{\ast}-\rho)^{-1}]^a_k\left(\mathring{\nabla}_i\rho^k_b-\rho^k_cT^c_{bi}\right)~,
\label{phigen}
\eea
where $\G^a_{bi}$ are the coefficients of the Levi-Civita connection $\nabla^{LC}$ on $D$, $\mathring{\nabla}$ is the Levi-Civita connection on $TM$ 
and 
\be \label{torsion}
T^a_{bi}=[(\theta^{\ast}+\rho)^{-1}]^a_k \left(\mathring\nabla_i(\theta^{\ast}+\rho)^k_b-\sfrac 12\rho^l_bH^k_{li}\right)~,
\ee
where one index of the 3-form $H$ is raised by means of the metric $g$. This is proven by direct computations. Indeed the 
first invariance condition
\eqref{cg} holds because
\bea 
\left({\cal L}_{\rho_a}g\right)_{ij}&=&
\rho^k_{a}\partial_kg_{ij}+2g_{k(i}\partial_{j)}\rho^k_{a}
\nn\\
&=& \rho^k_a\mathring\nabla_kg_{ij}+2\rho^k_a\G^l_{k(i}g_{j)l}+2g_{k(i}\mathring\nabla_{j)}\rho^k_a-
	2g_{k(i}\G^k_{j)l}\rho^l_a+2g_{k(i}\G^b_{j)a}\rho^k_b
\nn\\
&=& 2g_{k(i}\left(\mathring\nabla_{j)}\rho^k_a+\G^b_{j)a}\rho^k_b\right)
\nn\\
&=& 2g_{k(i}\left((\theta^{\ast}-\rho)^k_b\phi^b_{aj)}+\rho^k_bT^b_{aj)}+
	\rho^k_b\omega^b_{aj)}+\rho^k_b\phi^b_{aj)}-\rho^k_bT^b_{aj)} \right)
\nn\\
&=& 2g_{k(i}\left((\theta^{\ast})^k_b\phi^b_{aj)}+\rho^k_b\omega^b_{aj)} \right)
\nn\\
&=&\left( \omega^b_{a}\vee \iota_{\rho_b}g+\phi^b_{a}\vee 
\theta_b \right)_{ij}~,
\eea  
as required. The second condition is proven as shown below, for Lorentzian world sheets:
\bea 
 &&\left(\iota_{\rho_a}H -  \dd\theta_a+\omega^b_{a}\w\theta_b+\phi^b_a\w\iota_{\rho_b}g\right)_{ij}=
\nn\\
&\overset{\eqref{omegagen}}=&
\rho_a^kH_{kij}-2\partial_{[i}\theta_{aj]}+2(\G^b_{a[i}+T^b_{a[i})\theta_{bj]}-2\phi^b_{a[i}(\theta-\iota_{\rho} g)_{bj]}=
\nn\\
&\overset{\eqref{phigen}}=&\rho^k_aH_{kij}-2\mathring\nabla_{[i}(\theta^{\ast}+\rho)^k_ag_{\underline{k}j]}+2T^b_{a[i}\theta_{bj]}
+2\rho^k_cT^c_{a[i}g_{\underline{k}j]}=
\nn\\
&\overset{\eqref{torsion}}=&0~,
\eea
 as required. 

Having found the coefficients $\omega^a_{bi}$ and $\phi^a_{bi}$, it is now simple to write down the two connections $\nabla^{\pm}$. 
Let $T(\rho)=T^b_a\otimes e^a\otimes \rho_b \in \G(T^{\ast}M\otimes D^{\ast}\otimes TM)$ and denote 
by the same letters $(\theta^{\ast}\pm\rho)^{-1}\in \G(T^{\ast}M\otimes D)$ the sections of the indicated bundle 
that correspond to the operators $(\theta^{\ast}\pm\rho)^{-1}$. Then 
\bea 
\nabla^{\o}&=&\nabla^{LC}-\phi+T~,\\
\phi&=&\iota_{(\mathring\nabla-T)(\rho)}(\theta^{\ast}-\rho)^{-1}
~,
\eea
where the contraction is among the single $TM$ and $T^{\ast}M$ indices of the corresponding sections.
This leads directly to 
\begin{empheq}[box=\fbox]{align}\label{nablaplus}
 \nabla^+&=\nabla^{LC}+T~,\\
 \nabla^-&=\nabla^{LC}+T-2\iota_{(\mathring\nabla-T)(\rho)}(\theta^{\ast}-\rho)^{-1}~.\label{nablaminus}
\end{empheq}

We close this section by providing an alternative formulation of the main result. In particular,
having parametrized the Dirac structures in terms of orthogonal operators with the aid of the metric $g$, it is 
useful to express the connections $\nabla^{\pm}$  in terms of the operator ${\cal O}$ whose graph is $D$.
 Clearly this is more transparent in the parametrization in terms of the 
unconstrained field $\mf a=\mf a^i\partial_i\in \Omega^1(\S,X^{\ast}TM)$.
Its gauge transformation $\d_{\epsilon}\mf a^i=\dd\epsilon^i+C^i_{jk}\mf a^j\epsilon^k+\omega^i_{jk}\epsilon^j 
\dD X^k+\phi^i_{jk}\epsilon^j\ast \dD X^k$
 involves then the coefficients
\bea 
\omega^{i}_{jk}&=&\G^i_{jk} + \sfrac 12 ({\cal O}^{-1})_m^{\ \ i}\nabla_j{\cal O}^m_{\ k}+\sfrac 18(\text{id}+{\cal O}^{-1})_m^{\ \ i}H^m_{jl}(\text{id}-{\cal O})^l_{\ k}~,
\label{omegagen2} \\
\phi^i_{jk}&=&-\sfrac 12 ({\cal O}^{-1})_m^{\ \ i}\nabla_j{\cal O}^m_{\ k}+\sfrac 18(\text{id}-{\cal O}^{-1})_m^{\ \ i}H^m_{jl}(\text{id}-{\cal O})^l_{\ k}~.
\label{phigen2}
\eea
In coordinate independent form the connections $\nabla^{\pm}$ are
\bea
\nabla^+&=&\nabla^{LC}+\sfrac 12H\left(\text{id}-{\cal O}\right)~,\\
\nabla^-&=&\nabla^{LC}-\iota_{(\mathring\nabla-\sfrac 12 H)(\text{id}-{\cal O})}{\cal O}^{-1}~,
\eea
where $H(\text{id}-{\cal O})=\sfrac 12 H^i_{jk} \dd x^j \otimes (\text{id}-{\cal O})^l_{\ i}\partial_l \otimes\dd x^k
\in \G(T^{\ast}M\otimes TM\otimes T^{\ast}M)$ and $\co^{-1}\in\G(T^{\ast}M\otimes TM)$. The contracted indices are the $T^{\ast}M$ index of $\co$ with the $TM$ index of $(\mathring\nabla-\sfrac 12 H)(\text{id}-{\cal O})$. 
Then along with $\d_{\epsilon}X=(\text{id}-{\cal O})\epsilon$, the action functional \eqref{SDSM}
 is gauge invariant. As a final note, let us observe that in case $\phi=0$, namely when there is no term proportional to 
 $\ast \dD X^i$ in the gauge transformation of the gauge field, there is only one connection on $D$ and the result depends solely on 
 $\theta^{\ast}+\rho$.

\subsection{General form of the gauge symmetries for DSMs}
\label{dsmsymmetry}

Now we would like to finally determine the gauge symmetries for the DSM and 
show that they take the form suggested without proof in \cite{dsm1}. 
Note that we specialize here to the case $\ast^2=1$ as in the aforementioned paper.

Here we work in the parametrization of the Dirac structure in terms of the operator ${\cal O}$, using the unconstrained gauge field 
${\mf a}\in \Omega^1(\S,X^{\ast}TM)$ instead of the constrained ones $v\oplus \eta\in \Omega^1(\S,X^{\ast}D)$. Our purpose is then to 
determine $\d_{\epsilon}\mf a$. First of all, note that
\be \label{da1}
\d_{\epsilon}\mf{a}=\big(\d_{\epsilon}\mf{a}^i+\G^i_{jk}\mf{a}^j\d_{\epsilon}X^k\big)\partial_i=
\big(\d_{\epsilon}\mf{a}^i+\G^i_{jk}\mf{a}^j({\text{id}}-\co)^k_{\ l}\epsilon^l\big)\partial_i~,
\ee
since $\d_{\epsilon}X=(\text{id}-{\cal O})\epsilon$. Let us also recall that the gauge transformation for $\mf a^i$ is given as  
\bea 
\d_{\epsilon}\mf{a}^i&=&\dd\epsilon^i+C^i_{jk}(X)\mf{a}^j\epsilon^k+\D\mf a^i~,
\eea
where $\D\mf a^i=\omega^i_{jk}(X)\epsilon^j \dD X^k
+\phi^i_{jk}(X)\epsilon^j\ast \dD X^k$.
Two additional relations that will prove useful are the form of the pull-back $\mathring{\nabla}^{\ast}$ of the Levi-Civita connection 
$\mathring{\nabla}$ to $X^{\ast}TM$, 
\be
\mathring{\nabla}^{\ast}\epsilon =\left(\dd\epsilon^i+\G^i_{jk}\dd X^j\epsilon^k\right)\partial_i~,
\ee
and the expression of the structure functions $C^i_{jk}$ in terms of the operator ${\cal O}$, which is derived in the Appendix:
\be \label{dsmc}
 C^i_{jk}=
 2(\text{id}-\co)^l_{\ [j}\G^i_{k]l}+ g^{il}g_{mn}\co^n_{\ [j}\mathring\nabla_l\co^{m}_{\ k]}
 +\sfrac 12 H^i_{mn}(\text{id}-{\co})^m_{\ j}(\text{id}-\co)^n_{\ k}~,
 \ee
 where the weighted antisymmetrization is applied only on the indices $j$ and $k$.
Substituting the above in \eqref{da1} we 
obtain
\bea 
\d_{\epsilon}\mf{a}&=&\mathring\nabla^{\ast}\epsilon+\left(-\G^i_{jk}\dd X^j\epsilon^k+(\text{id}-\co)^l_{\ j}\G^i_{kl}\mf{a}^j\epsilon^k+\Delta\mf{a}^i\right)\partial_i+\nn\\
&&+\left( g^{il}g_{mn}(\co^n_{\ [j}\mathring\nabla_l\co^{m}_{\ k]}+\sfrac 12 g^{il}H_{lmn}\left(\text{id}-\co\right)^m_{\ j}\left(\text{id}-\co\right)^n_{\ k}\right)\mf a^j\epsilon^k\partial_i~.
\eea
Using that $\dD X^i=\dd X^i-(\text{id}-\co)^i_{\ j}\mf{a}^j$ and rearranging terms, we obtain
\be 
\d_{\epsilon}\mf{a}=\mathring\nabla^{\ast}\epsilon-g\left(\co^{-1}\mathring\nabla(\co)\mf{a},\epsilon\right)^{\ast}
+(\Delta\mf{a}^i-\G^i_{jk} \dD X^j\epsilon^k)\partial_i+\sfrac 12 H\left(\left(\text{id}-\co\right)\mf{a},
\left(\text{id}-\co\right)\epsilon,\cdot\right)^{\ast}~.
\ee
The last step amounts to substituting the expressions for $\omega^i_{jk}$ and $\phi^i_{jk}$ given in \eqref{omegagen2} and 
\eqref{phigen2} in $\D \mf a^i$. A little algebra leads to the final result for the general form of the gauge transformation 
of $\mf a$:
\begin{empheq} [box=\fbox]{align}
\d_{\epsilon} \mf a&=\mathring\nabla^{\ast}\epsilon-g\big(\co^{-1}\mathring\nabla(\co)\mf{a},\epsilon\big)^{\ast}
+\sfrac 12 H((\text{id}-{\cal O})\mf a,(\text{id}-{\cal O})\epsilon,\cdot)^{\ast}+\nn\\
& \quad +\Theta\left(\sfrac 12 H(\dD X,(\text{id}-{\cal O})\epsilon,\cdot)^{\ast}+(1-\ast)\mathring{\nabla}_{\dD X}({\cal O})\epsilon\right)~,
\label{dsmsymm}\end{empheq}
where we defined the operator
\be \label{Tinv}
\Theta=\sfrac 14 (\text{id}+{\cal O}^{-1})+\sfrac 14 (\text{id}-{\cal O}^{-1})\ast~
\ee
on $T^{\ast}\S\otimes X^{\ast}TM$. This operator is the inverse of the operator $(\text{id}+{\cal O})+(\text{id}-{\cal O})\ast$, 
denoted by $T$ in Ref. \cite{dsm1}. This is also proven in the Appendix.
Then \eqref{dsmsymm} is indeed the desired form for the gauge transformation.
Note that additional trivial gauge symmetries parametrized by an endomorphism 
$M\in \G(\text{End}(T^{\ast}\S\otimes X^{\ast}TM))$, as explained in \cite{dsm1},
may be included without any problem. 

\section{Application: Wess-Zumino Poisson $\sigma$-model}

The Poisson sigma model is a typical example in the class of topological DSMs, whose target space is a Poisson manifold.
An instructive way to obtain the model has ordinary 2D Yang-Mills theory as a starting point \cite{psm2}:
\bea\label{ym}
S_{\rm YM}[A]=\int_{\S}\sfrac 12 F\w\ast F~,\eea
where $F=(\dd A_i+ \sfrac 12 C^{jk}_iA_j\w A_k)e^i$ is the field strength of the gauge field $A=A_ie^i$ and $\{e^i\}$ are the 
generators of the gauge algebra. In the first order formalism we introduce conjugate variables $X^i$ to the gauge field $A_i$ 
and rewrite the action as
\bea\label{ym1}
S_{\rm YM}[X,A]=\int_{\S} \sfrac 12 \left(X^iF_i-\sfrac 12 (X^i)^2\ast 1\right)~.\eea
Inserting the field equation for $X^i$   into the action \eqref{ym1} reproduces the original action \eqref{ym} by construction.
Here, however, we are interested in the topological sector of the action \eqref{ym1}, i.e. omitting the potential term, 
which can be rewritten (on a world sheet without boundary and with a rescaling by a factor of 2)  as
\bea\label{top}
S_{\rm top}[X,A]=\int_{\S} \left(A_i \w \dd X^i+\sfrac 12 \pi^{ij}(X)A_i\w A_j\right)~. \eea
Taking this action as a starting point we interpret  $X^i$  as  pull-back coordinates on the target Poisson manifold $M$ with 
Poisson structure $\pi=\sfrac 12 \pi^{ij}(x)\partial_i\w\partial_j$ and the 
gauge fields $A_i$ as 1-forms on the world-sheet; thus we obtain the topological Poisson $\s$-model (PSM) \cite{psm1,psm2}.

In the following we are going to discuss the extension of the model that includes a WZ term \cite{wzpsm,wzpsm2}. 
The topological action corresponding to the WZPSM is 
\be\label{wzpsm}
S_{\text{WZPSM}}[X,A]=\int_{\S}\left(A_i\wedge \dd X^i + \sfrac{1}{2}\pi^{ij}(X)A_i\w A_j\right)+\int_{\hat\Sigma}H(X)~, 
\ee
for an antisymmetric 2-vector $\pi$ and a 3-form $H$.
 The symmetries of the model naturally  descend from the gauge symmetry of the Yang-Mills action \eqref{ym}. 
For  non-zero closed 3-form $H$ the action (\ref{wzpsm}) is invariant under
\bea\label{g1h}
\delta_{\epsilon}X^i &=& \pi^{ji}\epsilon_j~,\nn\\
\delta_{\epsilon}A_i &=& \dd\epsilon_i+\partial_i\pi^{jk}A_j\epsilon_{k}-\sfrac 12 H_{ijk}\pi^{kl}\epsilon_l(\dd X^j-\pi^{jm}A_m)~,\label{trs}
\eea
provided that the following condition 
\be \label{inth}
\pi^{il}\partial_l\pi^{jk}+\text{cycl}(ijk)=H_{i'j'k'}\pi^{i'i}\pi^{j'j}\pi^{k'k}~,
\ee
holds \cite{kss}.  
One immediately  recognizes this condition as the integrability condition for the Dirac structure given by
the graph of the 2-vector $\pi$, with the bracket on sections of $T^*M$ twisted by the closed
3-form $H$ \cite{wzpsm,wzpsm2}. Integrability conditions for more general Dirac structures were 
derived in \cite{cjl}.

In the spirit of \eqref{SDSM} we now add a kinetic  term  
and consider the action
\begin{equation}
S_{g\text{WZPSM}}[X,A] = \int_{\Sigma} \,\left( A_i \wedge \dd X^i + \sfrac{1}{2}\pi^{ij}A_i \wedge A_j\right)
+ \; \int_{\Sigma}\, 
\sfrac{1}{2}g_{ij}(X)\dD X^i\w\ast \dD X^j + \int_{\hat\Sigma} \, H(X) ~,\label{PSMg}
\end{equation}
with the covariant differential being
\be 
\dD X^i=\dd X^i+\pi^{ij}A_j~.
\ee
From the perspective of our approach, the relation to the general action for a DSM \eqref{SDSM}  
is established upon the identifications 
\be\label{ide} 
\eta_i= A_i \quad \text{and} \quad  v^i= \pi^{ji} A_j~.
\ee 
More specifically, the vector bundle $L$ is taken to be the cotangent bundle $L=T^{\ast}M$ and the maps 
$\rho: T^{\ast}M \to TM$ and $\theta: T^{\ast}M\to T^{\ast}M$ are 
\be \label{rhothetapsm}
\rho=\pi^{\sharp} \quad \text{and} \quad \theta=\text{id} ~ \Rightarrow ~ \theta^{\ast}=g^{-1}~,
\ee
with the former acting by contraction with $\pi$ on its first index.
The gauge field is valued in the (twisted) Lie algebroid  $(T^*M,[\cdot,\cdot]_{KS},\pi^\sharp)$, 
where the bracket is given by the $H$-twisted Koszul-Schouten bracket on 1-forms $\a,\widetilde \a$
\bea \label{kosH}
[\a,\widetilde\a]_{\text{KS}}:={\cal L}_{\pi^{\sharp}(\a)}\widetilde\a-\iota_{\pi^{\sharp}(\widetilde\a)}\dd \a
-H(\pi^{\sharp}(\widetilde\a),\pi^{\sharp}(\a),\cdot)~.
\eea
In a basis $\{e^i\}$ of local sections of $T^{\ast}M$ the bracket closes with a set of structure functions:
\be 
[e^i,e^j]_{\rm KS}=C^{ij}_k(X)e^k~,
\ee
this being \eqref{bracket} specialized in the present case.
These structure functions are found to be (cf. \cite{kss})
\be\label{CC}
C^{ij}_k=\partial_k\pi^{ij}+ H_{kmn}\pi^{mi}\pi^{nj}~.
\ee
Then the action \eqref{PSMg} 
is invariant under the infinitesimal gauge transformations
\begin{align}
\label{trafowz}
\delta_\epsilon X^i &= \pi^{ji} \epsilon_j \nonumber \\ 
\delta_\epsilon A_i &= \dd\epsilon_i +C^{jk}_{i}A_j\epsilon_k + \omega^j_{ik}\epsilon_j\dD X^k + \phi^j_{ik}\epsilon_j\ast \dD X^k~,
\end{align}
provided that   the  invariance conditions \eqref{cg} and \eqref{cH} and the constraints \eqref{cc1} and \eqref{cc2} hold. 
In particular the transformation of the gauge field takes the equivalent form
\bea
\label{tr}
\delta_{\epsilon}A_i = \dd\epsilon_i+\partial_i\pi^{jk} A_j\epsilon_k +H_{imk}\pi^{mn}\pi^{kl}A_n\epsilon_l+
\omega^j_{ik}\epsilon_j\dD X^k + \phi^j_{ik}\epsilon_j\ast \dD X^k~.\nn
\eea
Regarding the constraints, \eqref{cc1} is identically satisfied due to the antisymmetry of $\pi$, while 
\eqref{cc2} is also satisfied due to \eqref{CC}. 
As for the invariance conditions, the solution for the coefficients $\omega^i_{jk}$ and $\phi^i_{jk}$ can be read off 
the general formulae \eqref{omegagen}, \eqref{phigen} and \eqref{torsion}. Substitution of \eqref{rhothetapsm} in 
\eqref{torsion} directly gives
\be 
T^k_{ij}=(g^{-1}+\pi)^{-1}_{li}\left(\mathring\nabla_j\pi^{kl}-\sfrac 12 \pi^{km}H^l_{mj}\right)~.
\ee
Then \eqref{phigen} yields
\bea 
\phi^k_{ij}&=&(g^{-1}-\pi)^{-1}_{li}\left(\mathring\nabla_j\pi^{kl}-\pi^{ml}T^k_{mj}\right) \nn\\
	   &=&(g^{-1}-\pi)^{-1}_{li}\left(\mathring\nabla_j\pi^{kl}-\pi^{ml}(g^{-1}+\pi)^{-1}_{nm}\mathring\nabla_j\pi^{kn}+\sfrac 12
		\pi^{ml}\pi^{kp}(g^{-1}+\pi)^{-1}_{nm}H^n_{pj}\right) \nn\\
	   &=&(g^{-1}-\pi)^{-1}_{li}(g^{-1}+\pi)^{-1}_{nm}\left[\left((g^{-1}+\pi)^{ml}-\pi^{ml}\right)\mathring\nabla_j\pi^{kn}
	      +\sfrac 12\pi^{ml}\pi^{kp}H^n_{pj}\right] \nn\\
	   &=&(g^{-1}-\pi)^{-1}_{li}(g^{-1}+\pi)^{-1}_{nm}\left(g^{ml}\mathring\nabla_j\pi^{kn}+\sfrac 12\pi^{ml}\pi^{kp}H^n_{pj}\right)~\nn\\
&=&  (g^{-1}-\pi)^{-1}_{li}(g^{-1}+\pi)^{-1}_{nm}\left(g^{ml}(\mathring\nabla_j\pi^{kn}-\sfrac 12\pi^{kp}H^n_{pj})+\sfrac 12(g^{-1}+\p)^{ml}\pi^{kp}H^n_{pj}\right)~.\nn\\
\eea
Using the following relations:
\bea\label{aaa}
&&(g^{-1}-\pi)^{-1}_{li}(g^{-1}+\pi)^{-1}_{nm}~g^{ml}=[(1-g\pi g\pi)^{-1}]^k_ig_{kn}~,\\
&&(g^{-1}\pm\pi)^{-1}_{ik}=g_{ij}(1\mp\pi g)^j_l[(1-g\pi g\pi)^{-1}]^l_k~,\label{aaaa}
\eea
we can rewrite $\phi^k_{ij}$ as 
\be
\phi^k_{ij}=
 -[(1-g\pi g\pi)^{-1}]^l_ig_{lm}(\mathring\nabla_j\pi^{mk}+\sfrac 12 H_{jnp}\pi^{nm}\pi^{pk})~.
\ee
As for $\omega^{k}_{ij}$, from \eqref{omegagen} we find
\bea 
\omega^k_{ij}&=&\G^k_{ij}-\phi^k_{ij}+T^k_{ij} \nn\\
	      &=& 
	      \G^{k}_{ij}+[(1-g\pi g\pi)^{-1}]^l_ig_{lm}(\mathring\nabla_j\pi^{mk}+\sfrac 12 H_{jnp}\pi^{nm}\pi^{pk})
		 +\nn\\
		 && +(g^{-1}+\pi)^{-1}_{li}\left(\mathring\nabla_j\pi^{kl}-\sfrac 12 \pi^{km}H^l_{mj}\right)~. 
\eea
Again using the relation \eqref{aaaa} one obtains:
\be
 \o^k_{ij}=
 \Gamma^k_{ij}+g_{il}\pi^{lm}\phi^k_{mj}+\sfrac 12\pi^{kl}H_{lij}~.
\ee
Summarizing, the sought-for coefficients are found to be
 \bea\label{h0}
 \o^j_{ik}&=&
 \Gamma^j_{ik}+g_{il}\pi^{lm}\phi^j_{mk}+\sfrac 12\pi^{jl}H_{lik}~,
 \\
 \phi^j_{ik}&=&
 -[(1-g\pi g\pi)^{-1}]^l_ig_{lm}(\mathring\nabla_k\pi^{mj}+\sfrac 12 H_{knp}\pi^{nm}\pi^{pj})
 ~.\eea

Finally, let us mention for completeness that the relevant orthogonal operator ${\cal O}$ for the Dirac structure considered here 
has the form (cf. \cite{dsm1})
\bea\label{pO}
{\cal O}=(1-\pi g)(1+\pi g)^{-1}~,\eea
where $\pi g$ is understood as a map from $TM$ to itself acting as 
$\pi g(v^i\partial_i)=\pi^{ij}g_{jk}v^k\partial_i$. Then 
\bea \label{pOO}
\text{id}-\co&=&2\pi g(1+\pi g)^{-1}~,\\
\text{id}+\co&=&2(1+\pi g)^{-1}~,
\eea
and we can relate the gauge field $A\in \Omega^1(\S,X^{\ast}T^{\ast}M)$ to the gauge field 
$\mf a\in \Omega^1(\S,X^{\ast}TM)$ by means of \eqref{Aa}, which yields
\be 
\mf a^i=-\sfrac 12 (g^{-1}+\pi)^{ij}A_j~.
\ee

\section{Summary of results and conclusions}

Our main purpose in this work was to study the gauge symmetries of two-dimensional bosonic $\sigma$-models 
 beyond the standard approach. Thus we considered the bosonic string $\sigma$-model at leading order in 
 $\alpha'$ and asked under which conditions it can be extended by gauge fields $A\in \G(L)$, $L$ being a 
 Lie algebroid, such that it has a gauge symmetry $\d_{\epsilon}X^i=\rho^i_a(X)\epsilon^a(\s)$ 
 for a set of non-Abelian vector fields $\rho_a=\rho_a^i\partial_i$. These vector fields are obtained as the 
 image of some local basis of sections of $L$ under a homomorphism $\rho:L\to TM$. 
 The analysis, which will be presented in a more general context in \cite{cdjs}, leads to the following results:
 \begin{enumerate}
 \item In case the gauge fields are minimally-coupled to the theory, the gauging of the $\s$-model 
 is possible under the conditions \eqref{cg1} and \eqref{cB1}. These conditions relax the strict invariance 
 required in the ordinary case.
 \item In the presence of Wess-Zumino terms, the gauging of the $\s$-model is possible when the conditions 
 \eqref{cg} and \eqref{cH} supplemented by the constraints \eqref{cc1} and \eqref{cc2} hold. Once more, these conditions are 
 milder than in the ordinary gauging. Moreover, the 
 effective target space is the generalized tangent bundle $TM\oplus T^{\ast}M$.
 \end{enumerate} 
 Applying the above results led us to the main new results of the present paper: 
 \begin{enumerate}
 \item[3.] Geometrically, the gauging is controlled by two connections $\nabla^{\pm}$ 
  on the Lie algebroid $L$. In case $L$ and its image under the map $\rho\oplus\theta\colon L \to 
  TM\oplus T^{\ast}M$ are Dirac structures, namely maximally isotropic and involutive subbundles in the $H$-twisted standard Courant algebroid, these 
  connections are given by the formulae \eqref{nablaplus} and \eqref{nablaminus}. 
 \item[4.] The gauge symmetries of Dirac $\s$-models, constructed as general topological field theories based on Dirac structures, 
 take the general form \eqref{dsmsymm}.
\end{enumerate}   
Moreover these general results were applied to the case of the WZ-twisted Poisson $\s$-model with an 
auxiliary kinetic term, thus enabling us to determine the most general gauge symmetries for this model.

From these results one can conclude that two-dimensional $\sigma$-models can have richer gauge structure than one would expect, at 
least at the classical level. This can be useful either in studies of gauge theory in itself or in other contexts such as T-duality. 
Often the study of T-dual backgrounds is facilitated by gauge theory \cite{Buscher,Duff:1989tf}, however this requires 
background geometries
that contain isometries. The investigation of T-duality without isometries was initiated in Refs. \cite{cdj,cpos} in a  somewhat less 
general formulation than the one presented here. A challenging task is to examine whether this more general 
formulation can be applied to conformal string backgrounds that break isometries. 
Additionally, it is worth trying to understand the relation of our approach to 
Poisson-Lie T-duality, which is a technique that also does not require isometries \cite{pl1,pl2,pl3,pl4}. 
Furthermore, the geometric interpretation of our results reveals relations to generalized geometry. For example the 
effective target space of the gauged $\sigma$-models turns out to be the generalized tangent bundle $TM\oplus T^{\ast}M$. 
Extended targets for $\s$-models have been considered before, for example in 
\cite{Tseytlin:1990va,Siegel:1993th,Siegel:1993xq,Berman:2007xn,Berman:2007yf,Hull:2004in,Hull:2006qs} and more recently 
in \cite{Mylonas:2012pg,Bakas:2016nxt,Chatzistavrakidis:2015vka}. 
Notably, the relevance of the cotangent bundle was highlighted in \cite{Mylonas:2012pg} and also in 
\cite{Chatzistavrakidis:2015vka} in the context of Courant $\sigma$-models. 
It is interesting to explore further geometric aspects of such models through the gauging procedure employed 
here.
 
\paragraph{Acknowledgements.} 
A. Ch. would like to thank V. Penas, E. Plauschinn,  L. Romano, E. Saridakis and K. Sfetsos for discussions and suggestions. 
A.D. wants to thank J. Stasheff and D. Berman for discussion and the Erwin Schroedinger Institute in Vienna for hospitality.
We acknowledge  support by COST (European Cooperation in Science  and  Technology)  in  the  framework  of the  Action  MP1405  QSPACE. 
The work of L.J. and A. Ch. was supported  by Croatian Science Foundation under the project IP-2014-09-3258. 
L.J. was  supported by the H2020 CSA Twinning project No. 692194, "RBI-T-WINNING".

\appendix 

\section{A direct proof for the gauge symmetries of the PSM}
As an independent check of the results obtained in the main text,
in this appendix we provide a direct proof of the gauge symmetries \eqref{dsmsymm} for the purely Poisson sigma model,
i.e. with $H=0$. We consider maps $X: \Sigma \rightarrow M$ for $(M,\pi)$ a Poisson manifold. 
There is a canonical Lie algebroid structure $(T^*M,[\cdot,\cdot]_{KS},\pi^\sharp)$ with anchor given on standard basis 
elements by $\pi^\sharp(\dd x^i) = \pi^{ik}\partial_k$\footnote{To distinguish the coordinates on $M$ from the maps $X^i$,
we denote them by $x^i$.}. For convenience in reading we recall the gauging conditions in this case: 

\vspace{4pt}

{\bf Gauging conditions.} For $X:\Sigma \rightarrow M$ and $(M,\pi)$ a Poisson manifold, consider world sheet 1-forms 
$A_i \in \Omega^1(\S,X^*T^*M)$. Then the action of the Poisson sigma model 
\begin{equation}
S_{\text{PSM}}[X,A] =\;\int_\Sigma\,\Bigl(A_i \wedge \dd X^i + \tfrac{1}{2}\,\pi^{ij}A_i\wedge A_j\Bigr) +  \int_{\Sigma}\, 
\sfrac{1}{2}g_{ij}(X)\dD X^i\w\ast \dD X^j~,
\end{equation}
with $\dD X^i = \dd X^i - \pi^{ki}A_k$, is invariant under the transformation
\eqn{
\label{variation}
\delta_\epsilon X^i =&\,\pi^{ki}\epsilon_k \;,\\
\delta_\epsilon A_i =&\, \dd\epsilon_i + \partial_i \pi^{jk}A_j \epsilon_k + \Delta A_i\;,
}
if the following conditions on $M$ hold for $\Delta A_i =-\,\epsilon_k\omega^k_{ij} \dD X^j + \epsilon_k\phi^k_{ij} \ast \dD X^j$: 
\eqn{
\label{conditions}
{\cal L}_{\pi^\sharp \dd x^i}\,g =&\, \omega^i_j \vee \iota_{\pi^\sharp \dd x^j} g + \phi^i_j \vee \dd x^j\;, \\
0 =&\, -\omega^i_j \wedge \dd x^j \pm \phi^i_j \wedge \iota_{\pi^\sharp \dd x^j}\,g\;.
}

The goal of this appendix is to give a solution to the conditions \eqref{conditions} for $\phi$ and $\omega$, use them to 
rewrite the variation \eqref{variation} of $X$ and $A$ and confirm that they coincide with the form given in the literature 
(prop.7 of \cite{dsm1}) and also found in Section \ref{dsmsymmetry}, specialized to the case of vanishing WZ term.
Let us first recall that in the case of the PSM, the operator $\co$ is uniquely determined by the metric and Poisson tensor, \eqref{pO} and \eqref{pOO}. 
In particular the map $\pi g: TM\to TM$ is given as the Cayley transform of the operator ${\cal O}$ is given by
\begin{equation}
\bar{{\cal O}} =\; (1-\co)(1+\co)^{-1} =\;\pi g\;.
\end{equation}
With these preparations, we can give a solution to the conditions \eqref{conditions}, 
as they were already formulated in the main text, \eqref{h0}, and are used here for an independent proof: 
\eqn{
\label{solutionspsm}
\omega^i_{mn} =&\; \Gamma^i_{mn} +g_{mk}\pi^{kr}\phi^i_{rn}\;,\\
\phi^i_{mn} =&\; \Bigl[(1 -g\pi g\pi)^{-1}\Bigr]_n^k (\mathring\nabla_m \bar\co)^i{}_k\;.
}
We now use these solutions in the variation of the fields $X$ and $A$. To compare the variations with the literature, 
we first introduce $\epsilon^m$ as $\epsilon_k = -2(g^{-1}+\pi)^{-1}_{km} \epsilon^m$ and note that the 1-forms $A_i$ 
are related to the 1-forms ${\mf a}^i$ of \cite{dsm1} by ${\mf a}^k=-\tfrac{1}{2}(g^{-1}+\pi)^{kr}A_r$, i.e. the variation of the ${\mf a}^k$ is determined by
\begin{equation}
\label{vara}
-\delta_\epsilon {\mf a}^k = \;\half \partial_m\Bigl(g^{-1}+\pi\Bigr)^{kr}\delta_\epsilon X^m A_r + \half\Bigl(g^{-1}+\pi\Bigr)^{kr}\delta_\epsilon A_r\;.
\end{equation}
Expressing the variations of $X^i$ and $A_k$ in terms of $\epsilon^k$ and ${\mf a}^k$ together with the solutions for $\omega$ and $\phi$, we arrive at the following intermediate result:
\begin{obser}
The variations of the fields $X^i$ and ${\mf a}^k$ are given by
\eqn{
\label{var2}
\delta_\epsilon X^i =\,&(1-\co)^i{}_k \epsilon^k\;,\\
-\delta_\epsilon {\mf a}^k =\,&2\Bigl(-\pi^{mp}\partial_m(g^{-1}+\pi)^{kn} + (g^{-1}+\pi)^{km}\partial_m \pi^{np}\Bigr)(g^{-1}+\pi)^{-1}_{pq}\epsilon^q(g^{-1}+\pi)^{-1}_{nr}{\mf a}^r \\
&-\dd \epsilon^k -(g^{-1}+\pi)^{km}\partial_p(g^{-1}+\pi)^{-1}_{mn}\,\epsilon^n\,\dd X^p \\
&+ (g^{-1}+\pi)^{kr}(g^{-1}+\pi)^{-1}_{pm}\epsilon^m\Gamma^p_{rn}\dD X^n \\
&-(g^{-1}+\pi)^{kr}(g^{-1}+\pi)^{-1}_{pm}\epsilon^m(-g_{rq}\pi^{qs} + \delta^s_r\otimes \ast)\phi^p_{sn}\dD X^n\;.
}
\end{obser}
 We now transform the result of the previous observation, to be compared with \eqref{dsmsymm} and the result of \cite{dsm1}. We need the explicit form of the operator $\Theta$ given in \eqref{Tinv} for the case of PSM: 
\eqn{\Theta= \; \half (g^{-1}+\pi) g(1-\pi g\pi g)^{-1}(1-\pi g \otimes \ast)\;.}
 Furthermore, we use the fact that the derivative of the Cayley transform of $\co$ is given by
\begin{equation}
\mathring\nabla_n \bar \co =\; 2(1+\co)^{-1}(\mathring\nabla_n \co) (1+\co)^{-1}\;.
\end{equation}
With the help of the latter, we can write the last line of \eqref{var2} in the form
\eqn{
-\tfrac{1}{2}(g^{-1}+\pi)^{kr}g_{mq}\Bigl((\co^{-1}\mathring\nabla_n \co)(1-\bar\co)^{-1}\Bigr)^q{}_s(\textrm{id}+\bar\co -\textrm{id}+\textrm{id}\otimes \ast)^s{}_r\,\epsilon^m\,\dD X^n\;.
}
This expression has two terms. The one multiplying with $(\textrm{id}-\textrm{id}\otimes \ast)$ gives, 
after noting that $\co^t =\,g\co^{-1}g^{-1}$, due to the orthogonality of $\co$ with respect to the metric $g$,
the desired expression. The second term is contracting with $\dD X^n$ and $\epsilon^p$ and can be added to the third line
of \eqref{var2}. We arrive at 
\eqn{
\label{var3}
-\delta_\epsilon &{\mf a}^k=\,-\Theta(\textrm{id}-\textrm{id}\otimes \ast) \mathring\nabla_{\dD X}(\co)\epsilon \\
&+\tfrac{1}{2}\Bigl(-\pi^{mp}\partial_m(g^{-1}+\pi)^{kr} + (g^{-1}+\pi)^{km}\partial_m \pi^{rp}\Bigr)\Bigl(g(1+\co)\Bigr)_{pn}\epsilon^n \Bigl(g(1+\co)\Bigr)_{rq}{\mf a}^q\\
 &-\dd\epsilon^k-\Bigl((1+\co)^{-1}g^{-1}\Bigr)^{kr}\partial_p\Bigl(g(1+\co)\Bigr)_{rn}\, \epsilon^n\,\dd X^p\\
&+\Bigl((1+\co)^{-1} g^{-1}\Bigr)^{kr}\Bigl(g(1+\co)\Bigr)_{pn}\epsilon^n\,\Gamma^p_{rm}\,\dD X^m\;,\\
&-\Bigl((1+\co)^{-1}g^{-1}\Bigr)^{kr}\Bigl(\co^{-1}(\mathring\nabla_{\dD X}\co)(1-\bar\co)^{-1}(1+\bar\co)\Bigr)^q{}_r \,g_{qn}\,\epsilon^n\;.
}
As a next step, we observe that when replacing $\partial_p$ by $\mathring\nabla_p$ in the third line of \eqref{var3}, 
we have to subtract the two missing Christoffel terms, one of which is of the form $-\Gamma^k_{pn}\,\dd X^p\,\epsilon^n$ (i.e.
combines with $-\dd\epsilon^k$ to $-\mathring\nabla^*\epsilon^k$), and the other one adds to the fourth line of \eqref{var3}, 
transforming $\dD X^n$ to $\pi^{nm}A_m$ there. Furthermore, in the last line, we use the definition of the Cayley transform and 
orthogonality of $\co$ w.r.t. the metric $g$ to simplify. Finally, in the first line, we use that $\pi$ is Poisson, i.e.
$\pi^{im}\partial_m\pi^{jk} + \textrm{cycl}^{ijk}=0$ to simplify the terms with two Poisson tensors. We also note, that one 
of the terms with one metric and one Poisson tensor is precisely the term $g(\co^{-1}(\mathring\nabla\co){\mf a}, \epsilon)^*$, 
if we replace partial by Levi-Civita covariant derivatives and subtract the two corresponding Christoffel terms. We arrive at
\eqn{
\label{var4}
-\delta_\epsilon &{\mf a}^k =\,\Bigl(g(\co^{-1}(\mathring\nabla \co){\mf a},
\epsilon)^{\ast} -\Theta(\textrm{id}-\textrm{id}\otimes \ast)(\mathring\nabla_{\dD X}\co) \epsilon - \mathring\nabla^*\epsilon\Bigr)^k \\
&+\frac{1}{2}\Bigl(-g^{km}\Gamma^r_{mu}\pi^{up} -\pi^{rm}\partial_m\pi^{pk} + \pi^{pm}\partial_m g^{kr} - \pi^{km}\Gamma^p_{mu}\pi^{ur}\Bigr)\times \\
&\times \Bigl(g(1+\co)\Bigr)_{pn}\epsilon^n\Bigl(g(1+\co)\Bigr)_{rq}{\mf a}^q\\
&-\Bigl((1+\co)^{-1}\Bigr)^k{}_r \mathring\nabla_{\dd X}(1+\co)^r{}_n\epsilon^n + \Bigl((1+\co)^{-1}\Bigr)^k{}_r \mathring\nabla_{\dD X} \co^r{}_n\,\epsilon^n\;.
}   
Finally, writing out the last three lines in components, after simplifying we get the  contribution: 
$
\Gamma^k_{mn}\,(1-\co)^m{}_q\,\epsilon^q\,{\mf a}^n,
$
which is the term needed to complete the variation of the fields ${\mf a}^k\partial_k$, as we have the following variation from \cite{dsm1}:
\eqn{
\delta_\epsilon {\mf a}=\; (\delta_\epsilon {\mf a}^k + \Gamma^k_{mn}{\mf a}^n(1-\co)^m{}_n\epsilon^n)\partial_k ~.
}
Thus, as a result we have given a proof of the following 
\begin{prop}
The following variations lead to a symmetry of the Poisson sigma model without WZ term: 
\eqn{
\delta_\epsilon X =\;& (1-\co)\epsilon\;,\\
\delta_\epsilon {\mf a} =\;&\mathring\nabla^*\epsilon - g(\co^{-1}\mathring\nabla(\co) {\mf a}, \epsilon)^* + \Theta(\rm{id}-\rm{id}\otimes \ast)\nabla_{DX}(\co)\epsilon \;.
}
\end{prop}

\section{Structure functions for Dirac structures}

In this brief Appendix we collect some technical calculations related to the proof of the form of the gauge symmetries in the 
DSM case, as it appears in Section \ref{dsmsymmetry}.

We would like to prove the formula \eqref{dsmc} for the structure functions of an arbitrary Dirac 
structure $D$ parametrized with the orthogonal operator ${\cal O}$.
 First we recall that the bracket on the Dirac structure $D$ is given by the standard Courant bracket twisted  by the closed 3-form $H$, given by the formula \eqref{courant}. Local sections on $D$, parametrized as  
\be \label{dssection}
v\oplus \eta=\left(\left(\text{id}-\co\right)^i_{\ k}\mf a^k\partial_i+g_{ij}\left(\text{id}
+\co\right)^j_{\ k}\mf a^k\dd x^i\right):=\mf u_i\mf a^i~,
\ee 
 should close under the bracket:
\be \label{abb}
[\mf u_i,\mf u_j]=C_{ij}^k(X)\mf u_k~.
\ee 
 Combining \eqref{dssection} and \eqref{abb}, a long but straightforward calculation gives
 \bea\label{cA1}
 ({\rm id}-\co)^l_{\ i}C^i_{jk}&=&({\rm id}-\co)^m_{\ j} \partial_m({\rm id}-\co)^l_{\ k}-({\rm id}-\co)^m_{\ k} \partial_m({\rm id}-\co)^l_{\ j}~,
 \\
 ({\rm id}+\co)^l_{\ i}C^i_{jk}&= & \sfrac 12 g^{li}
 g_{mn}({\rm id}+\co)^n_{\ k}\partial_i
 ({\rm id}-\co)^{m}_{\ j}-\sfrac 12 g^{li}
 g_{mn}({\rm id}+\co)^n_{\ j}\partial_i
 ({\rm id}-\co)^{m}_{\ k}-\nn \\ 
 &&-\sfrac 12 g^{li} ({\rm id}-\co)^m_{\ j}\partial_i\big(g_{mn}({\rm id}+\co)^n_{\ k}\big)+\sfrac 12 g^{li} ({\rm id}-\co)^m_{\ j}
 \partial_i\big(g_{mn}({\rm id}+\co)^n_{\ k}\big)+\nn \\ 
 &&+g^{li}({\rm id}-\co)^m_{\ j}\partial_m\big(g_{in}({\rm id}+\co)^n_{\ k}\big)-g^{li}({\rm id}-\co)^m_{\ k}\partial_m\big(g_{in}({\rm id}+\co)^n_{\ j}\big)+\nn\\
&&+g^{li}H_{imn}({\rm id}-{\co})^m_{\ j}({\rm id}-\co)^n_{\ k}~.\label{cA2}
 \eea
 The structure functions can be determined from the obvious identity
 \be 
C^i_{jk}=\sfrac 12 ({\rm id}-\co)^i_{\ l}C^l_{jk}+\sfrac 12({\rm id}+\co)^i_{\ l}C^l_{jk}~.
\ee
This is facilitated by extending the partial derivatives to covariant derivatives and then collecting the appropriate terms.
First, we rewrite (\ref{cA1}) as
\bea
({\rm id}-\co)^l_{\ i}C^i_{jk}&=&-({\rm id}-\co)^m_{\ j}\mathring \nabla_m\co^l_{\ k}+({\rm id}-\co)^m_{\ k} \mathring\nabla_m\co^l_{\ j}\nn\\
&&+\co^n_{\ k}({\rm id}-\co)^l_{\ i}
\Gamma^i_{jm}-\co^n_{\ j}({\rm id}-\co)^l_{\ i}\Gamma^i_{km}~.
\label{cc11}\eea
Similarly, Eq. (\ref{cA2}) may be rewritten as
\bea\label{cc21}
({\rm id}+\co)^l_{\ i}C^i_{jk}&= &-\sfrac 12 g^{li}
g_{mn}({\rm id}+\co)^n_{\ [k}(\mathring\nabla_i\co^{m}_{\ j]}-\co^p_{\ j]}\Gamma^m_{pi}+\co^m_{\ p}\Gamma^p_{j]i})-\nn\\
&-&\sfrac 12 g^{li}({\rm id}-\co)^m_{\ [j}\big(g_{mn}\mathring\nabla_i\co^n_{\ k]}+g_{pn}({\rm id}+\co)^n_{\ k]}\Gamma^p_{mi}+g_{mn}({\rm id}+\co)^n_{\ p}
\Gamma^p_{k]i}\big)
+\nn\\
&+&({\rm id}-\co)^m_{\ [j}\mathring\nabla_m\co^l_{\ k]}+H^l_{mn}({\rm id}-{\co})^m_{\ j}({\rm id}-\co)^n_{\ k}+\nn\\
&+&g^{li}({\rm id}-\co)^m_{\ [j}\big(g_{pn}({\rm id}+\co)^n_{\ k]}\Gamma^p_{im}+g_{in}({\rm id}+\co)^n_{\ p}
\Gamma^p_{k]m}\big)~.
\eea
Now for the first term in the third line we substitute the corresponding expression from (\ref{cc11}). 
Then several terms cancel and we obtain
\bea\label{cca}
C^l_{jk}=g^{li}g_{mn}\co^n_{\ [j}\mathring\nabla_i\co^m_{\ k]}-(\co^i_{\ j}\Gamma^l_{ki}-\co^i_{\ k}\Gamma^l_{ji})+\sfrac 12 H^l_{mn}({\rm id}-{\co})^m_{\ j}({\rm id}-\co)^n_{\ k}~,
\eea
and it is easy to see that this is the same as the desired expression \eqref{dsmc}.

\paragraph{Inverse of the operator $(\text{id}+{\cal O})+(\text{id}-{\cal O})\ast$~.} 
A final technical point concerns the inverse of the operator $(\text{id}+{\cal O})+(\text{id}-{\cal O})\ast$, 
which appears in the general form of the gauge symmetries for the DSM in \eqref{dsmsymm}. As discussed there, for Lorentzian world sheets this operator is 
\be 
\Theta=\sfrac 14(\text{id}+{\cal O}^{-1})+\sfrac 14 (\text{id}-{\cal O}^{-1})\ast~.
\ee
Indeed, using $\ast^2=1$, the following calculation establishes the statement:
\begin{align}
&\left((\text{id}+{\cal O})+(\text{id}-{\cal O})\ast\right)\circ \left(\sfrac 14(\text{id}+{\cal O}^{-1})+\sfrac 14 (\text{id}-{\cal O}^{-1})\ast\right)=\nn\\
&=\sfrac 14 (\text{id}+{\cal O})\circ (\text{id}+{\cal O}^{-1})+\sfrac 14 (\text{id}-{\cal O})
\circ(\text{id}-{\cal O}^{-1})+\nn\\
&+\sfrac 14(\text{id}+{\cal O})\circ(\text{id}-{\cal O}^{-1})\ast+\sfrac 14 (\text{id}-{\cal O})\circ(\text{id}+{\cal O}^{-1})\ast=\nn\\
& =\sfrac 14\left(2+{\cal O}+{\cal O}^{-1}\right)+\sfrac 14 \left(2-{\cal O}-{\cal O}^{-1}\right)+\sfrac 14 \left({\cal O}-{\cal O}^{-1}\right)\ast+\sfrac 14 \left({\cal O}^{-1}-{\cal O}\right)\ast=1~,
\end{align}
as required.

\end{document}